\documentstyle[prd,aps,preprint,floats,epsfig]{revtex}

\tighten

\begin{document}
\draft
 
\pagestyle{empty}

\preprint{
\noindent
\hfill
\begin{minipage}[t]{3in}
\begin{flushright}
UH-511-1034-03\\
LBNL--53823 \\
September 2003
\end{flushright}
\end{minipage}
}

\title{On the hidden charm state at 3872 MeV}

\author{ 
Sandip Pakvasa$^a$ and Mahiko Suzuki$^{b,c}$
}
\address{
$^a$Department of Physics and Astronomy, University of Hawaii, Honolulu, 
Hawaii 96822\\
$^b$Department of Physics and Lawrence Berkeley National Laboratory\\
$^c$University of California, Berkeley, California 94720
}


\date{\today}
\maketitle

\begin{abstract}
  We discuss some puzzling aspects of the narrow hidden charm resonance 
that was recently discovered by the Belle Collaboration at mass 3872 MeV.
In order to determine its quantum numbers, a crucial piece of information 
is the spin of the dipion in the decay final state $\pi^+\pi^-J/\psi$.
We give the angular distributions and correlations of the final 
particles in the decay which will provide this information 
about the nature of this resonance.

\end{abstract}
\pagestyle{plain}
\narrowtext

\setcounter{footnote}{0}

\section{Introduction}

A new narrow resonance has been discovered at mass 3872 MeV by the Belle
Collaboration\cite{Belle1} through the process,
\begin{equation}
B^{\pm}\to K^{\pm}X(3872)\to K^{\pm} \pi^+\pi^- J/\psi.
                                    \label{prod}
\end{equation}
Following Belle\cite{Belle2}, we denote this resonance
tentatively by $X(3872)$ in this paper.
The invariant mass of $\pi^+\pi^-$ extends to the upper end of its 
kinematical boundary ($\simeq$ 775 MeV), but it is not 
known at present whether the dipion is in $s$-wave, $p$-wave, 
or even $d$-wave. Experiment does not exclude the possibility 
that the dipion in the $\rho$ mass region is actually in $s$-wave. 
We should keep in mind that 
the $s$-wave $\pi\pi$ scattering cross section rises rapidly starting 
just below the $\rho$ mass. In fact, this partial-wave or spin of 
the dipion provides the most important clue to the quantum numbers 
of this resonance. Search of the radiative decays $X(3872)\to\gamma
\chi_{c1}$ has so far not produced a positive result\cite{Belle2}. 
As the upper bound set on the raditative decay branching becomes
more severe, it will impose a strong constraint on the nature of 
the resonance. 

The most likely candidate for $X(3872)$ is an excited charmoniun
state. In fact, the experimental study of the $\pi^+\pi^-J/\psi$ 
final state in $B$ decay was motivated with search of the excited 
charmonia\cite{Eich}. 
Meanwhile, closeness of the $D\overline{D}^*$ threshold to 3872 MeV 
suggests another explanation, a loosely bound molecular state 
of $D\overline{D}^*$ and $\overline{D}D^*$\cite{Georgi}. However,  
both the charmonium and the molecule interpretation may encounter 
some problems. In this paper we are not rigidly constrained by 
the potential model predictions on charmonia since uncertainties 
are large for the excited charmonium states near the open charm 
thresholds. We will be open minded about dynamics of the molecules.
Instead we narrow down the possible quantum numbers of $X(3872)$ 
with the information extracted from the Belle experiment and 
then focus on possibility of determining the quantum numbers 
purely experimentally by the decay angular distributions and 
correlations.

\section{Current experimental information}

The Belle experiment\cite{Belle1} has so far imposed the following 
constraints on this resonance. The
width is narrow ($<$ 2.3 MeV\cite{Belle2}) despite the ample 
phase space ($p_{cm}\simeq 500$ MeV) for the decay 
$X(3872)\to D\overline{D}$. For comparison, the width of 
$\psi''(3770)\to D\overline{D}$ is 24 MeV with $p_{cm}=242$ MeV. 
Provided that the decay into $D\overline{D}$ is forbidden 
by selection rules of quantum numbers rather than by some 
unknown dynamical suppression, we expect that $X(3872)$ should 
have an unnatural spin-parity,
\begin{equation} 
     J^{P}=0^-, 1^+, 2^-\cdots, 
\end{equation} or an unnatural spin-charge-parity,
\begin{equation}
     J^{PC}= 0^{+-}, 1^{-+}, 2^{+-}\cdots.
\end{equation}
To select a right $J^{PC}$ out of these choices, 
the dipion quantum numbers provide the most 
important clue. If $\pi^+\pi^-$ forms a scalar dipion of 
$J^{PC}_{\pi\pi}=0^{++}$ or a tensor dipion of $J^{PC}_{\pi\pi}=2^{++}$,
charge parity $C$ of $\pi^+\pi^-J/\psi$ is negative and isospin is $I=0$ 
or 2. We do not consider $I=2$ further in view  of lack of any candidate. 
Combining $\pi^+\pi^-$ with $J/\psi$ in 
relative orbital angular momentum $L$, one can make the unnatural spin 
parity and charge-parity states of $0^{+-}$, $1^{+-}$, $2^{\pm-}\cdots$
with a scalar or a tensor dipion. 
If $\pi^+\pi^-$ forms a vector dipion of $J^{PC}_{\pi\pi}=1^{--}$, 
charge parity of $\pi^+\pi^-J/\psi$ is positive and isospin is $I=1$. 
The most relevant unnatural spin-parity state is $1^{++}$ with $L=0$ 
in this case. These unnatural quantum number states are listed 
in Table I. Since only a limited phase space is available for the 
$\pi^+\pi^-J/\psi$ decay, the Table includes only the cases of
$L\leq 2$ for the scalar dipion, and $L\leq 1$ for the vector 
and tensor dipion.

\section{Charmonia}

Since $I=0$ for the charmonia, the folllowing quantum numbers are 
selected for charmonium candidates:
\begin{equation}
     J^{PC}= \left\{ \begin{array}{ll}
                  2^{--}, & (L=0,2)\\
                  1^{+-}, & (L=1).
             \end{array} \right.
\end{equation}
The charmonia carrying these quantum numbers are 
$^3D_2(2^{--})$ and $^1P_1(1^{+-})$.

The mass spectrum calculations by potential models\cite{Pot,Isgur} 
suggest that the $^3D_2$ state should be much closer to the
$1^3D_1$ state, which is believed to be $\psi^{''}(3770)$. 
The mass of the $1^3D_2$ state was predicted at 30$\sim$60 MeV lower 
than 3872 MeV\cite{Pot}. However, the charmonium levels near the 
$D\overline{D}$ and $D\overline{D}^*$ thresholds are subject to 
large uncertainties due to the open charm channel coupling and 
the large relativistic corrections ($\sqrt{\langle v^2/c^2\rangle} 
\simeq 0.54$ in one estimate\cite{Isgur}). Therefore we are not 
too concerned with the discrepancy between experiment and 
the potential model expectation for the masses in the case of those 
higher excited charmonia.

 The radiative decay into $\gamma\chi_{cJ}$ is allowed for both
$2^{--}$ and $1^{+-}$ by quantum numbers. The decay 
$^3D_2(2^{--})\to\gamma\chi_{cJ}$ is the $E1$ transitions and
should occur without suppression. Let us compare it with 
a similar radiative
decay of $\psi(2S)$.  We know $\Gamma(\psi(2S)\to\chi_{c1})
\simeq 26$ keV and $\Gamma(\psi(2S)\to \pi^+\pi^-J/\psi)\simeq 93$
MeV from experiment. The strong decay $\psi(2S)\to\pi^+\pi^-J/\psi$ 
occurs into a scalar dipion with $s$-wave ($L=0$) relative to $J/\psi$. 
Apart from difference in the dipole strength $\langle r\rangle$ 
between $\psi(2S)\to\chi_{cJ}$ and $1^3D_2\to\chi_{cJ}$, the spin 
factors and phase space corrections combined enhances 
$\Gamma(^3D_2(3872)\to\gamma\chi_{c1})$ by a factor of 
$\approx 5$ relative to $\Gamma(\psi(2S)\to\gamma\chi_{c1})$.
The dipole matrix element is sensitive to the coupling to the
$p$-wave $D\overline{D}^*$ channel. Comparison of the strong decay 
is at least as sensitive or even more uncertain since the decay $X(3872)\to 
\pi^+\pi^-J/\psi$ involves $d$ wave either in the dipion or in
the relative orbital angular momentum $L$. The rescaling factor
of the three-body phase space can be as large as $\approx 10$, 
which is severely compensated by the $d$-wave suppresion factor, 
$|{\bf p}_{\pi\pi}/E_{\pi\pi}|^4$ for a scalar dipion and 
$|{\bf p}_{\pi}/E_{\pi}|^4$ for a tensor dipion. It depends 
sensitively on the dipion mass distribution. Despite the 
larger phase space for $X(3872)\to\pi^+\pi^-J/\psi$ than 
for $\psi(2S)\to\pi^+\pi^-J/\psi$, the net result is
most likely that $\Gamma(X(3872)\to\pi^+\pi^-J/\psi)$ is 
smaller than $\Gamma(\psi(2S)\to\pi^+\pi^-J/\psi)$. With 
this order of magnitude consideration we expect  
$B(X(3872)\to\gamma\chi_{c1})/B(X(3872)\to\pi^+\pi^-J/\psi)> 1$.
Then the non-observation of the 
radiative decay $X(3872)\to\gamma\chi_{c1}$\cite{Belle2},
\begin{equation}
  B(X(3872)\to\gamma\chi_{c1})/B(X(3872)\to\pi^+\pi^-J/\psi)
         < 0.89 \;\;(90\% \;C.L.), \label{UB}
\end{equation}
poses a problem on the $^3D_2$ assignment. The Belle Collaboration
\cite{Belle2} compares the upper bound of Eq.(\ref{UB}) above 
with the potential model prediction of 5\cite{Eich}. Considering
the large uncertainties involved in the potential model
calculations, however, it is prudent to take a ``wait and
see'' attitude. 

The other possibility of the radially excited $^1P_1(1^{+-})$
encounters a larger discrepancy with the mass prediction of the
potential models. The mass 3872 MeV is roughly 100 MeV lower than 
the potential model. Nonetheless we do not reject $2^1P_1$ 
at present for the same reason stated above for $1^3D_2$. 
The radiative decay $2^1P_1\to\gamma\chi_{cJ}$ is an $M1$
transition that is caused by spin flip. Since the spatial wave
functions are orthogonal between $\chi_{cJ}(1^3P_J)$ and $2^1P_1$ 
in the nonerelativistic limit, the $M1$ transition is 
considerably weaker than the allowed $E1$ transitions. The strong 
decay $2^1P_1\to\pi^+\pi^-J/\psi$ occurs into a scalar dipion with
$L=1$ instead of $L=2$ for $1^3D_2\to\pi^+\pi^-J/\psi$. 
Therefore the radiative branching fraction is much smaller, 
relative to the $\pi^+\pi^-J/\psi$ decay branching, 
for $2^1P_1$ than for $1^3D_2$. 
It can be easily consistent with the current upper bound. If the
non-observation of the radiative decays into $\gamma\chi_{cJ}$ 
continues to a higher 
precision, we should consider $2^1P_1$ as a better candidate
than $1^3D_2$. We should keep in mind, however, that 
even the lowest $1^1P_1$ charmonium $h_c$ has not yet been 
reported in $B$ decay despite continuing searches by both the Belle 
and the BaBar Collaboration.

  To wit, if $X(3872)$ is a charmonium, it should be either
the $1^3D_2 (2^{--})$ state or the $2^1P_1(1^{+-})$ state.

\section{Molecules}

The idea of loosely bound molecule states of two hadrons
has been entertained for a long time\cite{Georgi}, but no meson has 
so far been positively identified as such a state. The close 
proximity of the mass 3872 MeV to the $D\overline{D}^*$ 
thresholds (3971.2 $\pm$ 0.7 MeV for $D^0\overline{D}^{*0}$ 
and 3979.3 $\pm$ 0.7 MeV for $D^+\overline{D}^{*-}$) has prompted 
reconsideration of this possibility for $X(3872)$\cite{Torn1}:
\begin{equation}
   (1/\sqrt{2})(D\overline{D}^* \pm \overline{D}D^*).
\end{equation}
Hereafter we refer to the charge-parity eigenstates of the molecule
states simply by $D\overline{D}^*$. Then the following unnatural
spin-parity and charge-parity states can be formed as a molecule:
\begin{equation}
         J^{PC}= \left\{ \begin{array}{ll}
               1^{+\pm} & (L_{D\overline{D}^*}=0),\\
               0^{-\pm}, 1^{-+}, 2^{-\pm} &(L_{D\overline{D}^*}=1).
         \end{array} \right. \label{molecules}
\end{equation} 
Since charge parity of $\pi^+\pi^-J/\psi$ is correlated to isospin 
by $C = -(-1)^I$, the molecule should be in $I=0$ for $C=-$ 
and in $I=1$ for $C=+$. For the relative orbital angular moementum
of $D\overline{D}^*$, the closeness of 3872 MeV to the 
$D\overline{D}^*$ threshold strongly favors $L_{D\overline{D}^*}=0$ 
over $L_{D\overline{D}^*}=1$. A qualitative argument of dynamics 
based on the one-pion exchange favors formation of the $I=0$ molecules 
over the $I=1$ ones, in particular, $0^{-+}$ and $1^{++}$\cite{Torn2}. 
However, the states of $0^{-+}$ and $1^{++}$ with $I=0$ are not 
found in Table I since they are inconsistent with the decay into
$\pi^+\pi^- J/\psi$. Is it possible that the states of $0^{-+}$ 
and $1^{++}$ with $I=0$ decay into the $I=1$ channels of 
$\pi^+\pi^-J/\psi$ with isospin symmetry breaking.
The answer is affirmative, but such a state cannot feed a scalar 
dipion in $\pi^+\pi^-J/\psi$ because of $C$ invariance.  In this 
case, the dipions observed with mass below the $\rho$ region 
would be entirely due to the vector dipion produced by
isospin breaking.
   
We do expect a large isospin violation in a molecular 
bound state of $D\overline{D}^*$ at 3872 MeV, since the 
mass 3872 MeV almost coincides with the $D^0\overline{D}^{^*0}$ 
threshold but as much as 7 MeV below the $D^+\overline{D}^{^*-}$ 
threshold. No matter how large isospin violation is, however,   
no single resonance can produce both a scalar dipion and a vector 
dipion in $\pi^+\pi^-J/\psi$ by $C$ invariance:
Any strong interaction resonance of 
zero net flavor must be an eigenstate of charge conjugation 
even when isospin is broken. Since the states 
of $I=0$ and $I=1$ have opposite $C$, a single resonance 
cannot be responsible for both the $I=0$ final state 
(a scalar dipion) and the $I=1$ final state (a vector dipion)
of $\pi^+\pi^-J/\psi$. 
In order to feed both the scalar/tensor and the vector dipion, 
there must be two (almost perfectly degenerate) 
resonances of opposite charge parities, one with $I=0$ 
and odd $C$ and the other with $I=1$ and even $C$.  Forming two
such molecules in degeneracy in hidden flavor channels 
would be a highly unlikely accident. 
With these observations, we should consider only a molecule of 
either $J^{PC}=1^{+-}$ with $I=0$ or $J^{PC}=1^{++}$ with $I=1$
as the molecule candidates among the entries in Eq. (\ref{molecules}).

It is likely that the branching fractions of the radiative transitions 
to $\gamma\chi_{cJ}$ are naturally small for the $D\overline{D}^*$
molecules in general. For $J^{PC}=1^{++}$, the radiative transitions
into $\gamma\chi_{cJ}$ is completely forbidden by $C$ invariance. 
Even for the molecules of $1^{+-}$, the radiative decay width 
should be rather small. Hence a small or negligible radiative decay 
width poses no problem for the interpretation in terms of 
a molecule. 

   If $J^{PC}$ of the most attractive channel of $D\overline{D}^*$
coincides with those of a charmonium, 
a substantial mixing can occur between $D\overline{D}^*$ and 
the charmonium\cite{B}. This is a real possibility for our most favorite 
charmonium candidates with $J^{PC}= 2{--}$ and $1^{+-}$.
If $X(3872)$ is such a mixed state of a molecule
and a charmonium, its radiative decay width is suppressed relative
to that of a pure charmonium by the fraction of the molecule 
mixing. 

   In fact, this kind of mixing is needed for a very loosely bound
molecule state to be produced in $B$ decay with a significant rate\cite{S}.
The production amplitude of a bound state is proportional to 
the ``wave function at origin'' $\Psi(0)$ for the same reason that  
the produciton amplitudes of $\pi$ and $K$ are proportonal to 
$f_{\pi}$ and $f_{K}$, respectively, which are interpreted as the
wave functons at the origin of quark-antiquark. For a loosely bound 
state with binding energy $\Delta E$, $|\Psi(0)|^2$ is a small 
quantity proportional to $(m\Delta E)^{3/2}$. For the $D\overline{D}^*$
molecule, this is much smaller than $|\Psi(0)|^2$ of
the charmonia for which the charm quark mass is not sharply defined.
If one accepts this argument, production of a pure $D\overline{D}^*$
molecule is minuscule relative to charmonium produciton in $B$ decay. 
Physically speaking, $D$ and $\overline{D}^*$ must fly in parallel with 
practically zero relative velocities in order to form a molecule
in $B$ decay. Such a phase space is a tiny, almost negligible 
fraction of the final-state phase space of $B$ decay. The only 
way to enhance the molecule production in $B$ decay is through a 
substantial mixing with a charmomium.\footnote{
For production of a pure $D\overline{D}^*$ molecule, a more favorable 
environment is near the $D\overline{D}^*$ threshold in $e^+e^-$ 
annihilation where the relative motion of $D$ and $\overline{D}^*$ 
is restricted to be small.}    
  
\section{Dipion and angular distributions}   

\subsection{Scalar dipion}
   The most important information in determining the quantum numbers
of $X(3872)$ is the spin of the dipion in $\pi^+\pi^-J/\psi$. Let us 
first consider the case that the dipions of $I=0$
are entirely in $s$-wave, $J^{PC}_{\pi\pi}=0^{++}$. If one recalls that
the d-wave $\pi\pi$ cross section is negligibly small in this region,
this is a reasonable possibility. However, it is experiment that should  
eventually determine 
whether $J_{\pi^+\pi^-}=0$ or not. If $\pi^+\pi^-$ is really a scalar 
dipion, the $\pi^+$ (or $\pi^-$) momentum in dipion rest frame should
show no angular correlation with other vectors;
\begin{equation}
    d\Gamma/d\Omega_{\pi^{\pm}}=\frac{1}{4\pi}\Gamma_0,
\end{equation} 
where $\Omega_{\pi^{\pm}}$ is the solid angle of ${\bf p}_{\pi^+}-
{\bf p}_{\pi^-}$ in the
dipion rest frame, as measured with respect to the direction of
any momentum, e.g. the $J/\psi$ momentum in the $X$ rest frame.
If this test shows that the dipion is indeed a scalar, the dipion 
angular distribution in the rest frame of $X(3872)$ 
happens to be independent of dynamics and unique for 
$J^{P}=1^{+-}$ and $2^{--}$ since the zero-helicity amplitude of 
$J/\psi$ is forbidden for them. We elaborate on this below. 

The $X(3872)$ state is produced in the zero-helicity state in the
$B$ rest frame when it is produced in $B^{\pm}\to K^{\pm}X(3872)$.
In the $X$ rest frame, $X(3872)$ is in $|J,0\rangle$ when
the quantization axis (call it the $z$-axis) of ${\bf J}$ is 
chosen along ${\bf p}_{X}$ in the $B$ rest frame.
According to the parity constraint of the helicity formalism\cite{JW}, 
the helicity amplitudes $\langle s_b,\lambda_b;s_c,\lambda_c|J_a,M\rangle$ 
for $a\to b+c$ in the $a$-rest frame obey
\begin{equation}
  \langle s_b,\lambda_b;s_c,\lambda_c|J,0\rangle =
   \eta_a\eta_b\eta_c(-1)^{J+s_b-\lambda_b+s_c-\lambda_c}
    \langle s_b,-\lambda_b;s_c,-\lambda_c|J,0\rangle,
\end{equation}
where $\eta_{a,b,c}$ are the instrinsic parities of $a(=X),
b(=\pi^+\pi^-),c(=J/\psi)$. In the present case, 
$\eta_b= (-1)^{J_{\pi\pi}}$ and $\eta_c (=\eta_{J/\psi})=-1$, and
$s_c=J_{J/\psi}=1$. 
We apply this relation to the case of a scalar dipion, 
$s_b=\lambda_b=0$. Denoting the helicity of $J/\psi$ by $h$ 
instead of $\lambda_c$, we obtain from the parity constraint
\begin{equation}
    \langle 0,0;1,h|J,0\rangle = \eta_a(-1)^{J-h}
                     \langle 0,0;1,-h|J,0\rangle .
\end{equation}
The factor $\eta_a(-1)^{J}$ is equal to $-1$ for $J^{P}=2^{-}$ 
and $1^{+}$. Therefore the $h=0$ amplitudes of $J/\psi$ must vanish:
\begin{equation}
   \langle 0,0;1,0|J,0\rangle = 0, \;\;
      (J^{PC}=2^{--} {\rm and}\; 1^{+-}). \label{zero}
\end{equation}
Let us suppose that one measures 
the angular distribution of the dipion momentum 
${\bf p}_{\pi\pi}={\bf p}_{\pi^+}+{\bf p}_{\pi^-}$ in the 
$X$ rest frame, choosing the $z$-axis along the 
$X$ momentum in the $B^{\pm}$ rest frame.
Following the standard formulae\cite{JW}, we obtain with Eq. 
(\ref{zero})  
\begin{equation}
  \frac{d\Gamma}{d\cos\theta_{\pi\pi}} \propto
              \left \{ \begin{array}{ll}
              \cos^2\theta_{\pi\pi}\sin^2\theta_{\pi\pi}, & (J^{PC}=2^{--})\\
              \sin^2\theta_{\pi\pi}, & (J^{PC}=1^{+-}),
                 \end{array} \right.
 \end{equation}
where $\theta_{\pi\pi}$ is the polar angle of ${\bf p}_{\pi\pi}$.
(See Fig.1.)
\begin{figure}[h]
\hspace*{5cm}
\epsfig{file=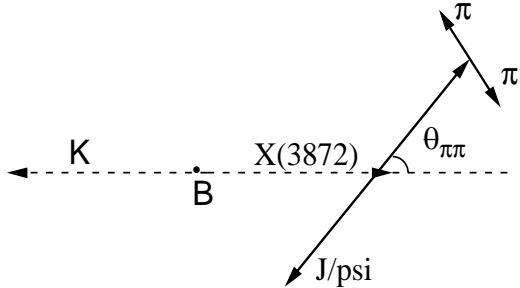,width=7cm,height=4cm}
\caption{The dipion scattering angle $\theta_{\pi\pi}$, 
which is defined in the $X$ rest frame.
\label{fig:1}}
\end{figure}

For the positive parity states other than $J^{PC}=1^{+-}$,
the longitudinal polarization decay amplitude enters the
angular distribution:
\begin{equation}
  \frac{{\rm d}\Gamma}{{\rm d}\cos\theta_{\pi\pi}} \propto
            \left \{ \begin{array}{ll}
                     1 & (J^{PC}=0^{+-}), \\
      \cos^2\theta_{\pi\pi}\sin^2\theta_{\pi\pi}
   +\frac{1}{6}\kappa(3\cos^2\theta_{\pi\pi}-1)^2
                       & (J^{PC}=2^{+-}),
           \end{array} \right.
\end{equation}
where $\kappa = |A_0|^2/(|A_1|^2+ |A_{-1}|^2)$  with $|A_1|= |A_{-1}|$.
The angular distribution for $J^{PC}=2^{+-}$ can mimic that of $2^{--}$
when $\kappa$ is small.

\subsection{Tensor dipion}

  The zero total helicity state $\lambda_{\pi\pi}-h=0$ 
can be produced as well as the nonzero helicity states for the 
tensor dipion. That is, more than one value of $L$ is allowed 
for a given $J^P$ of $X(3872)$ in the case of the tensor dipion.
Consequently the angular 
distributioins and correlations are dependent on dynamics. 
However, kinematics of the $X(3872)$ decay allow us 
to make a special approximation. Since the tensor dipion 
mass is produced with the invariant mass close to the upper 
end of the kinematical 
boundary ($\simeq 775$ MeV), the dipion and $J/\psi$ are 
most often produced nearly at rest in the $X$ rest frame. 
Let us select those fat dipions. We then expect that the 
relative orbital angular momentum between the dipion and 
$J/\psi$ is in s-wave ($L=0$) 
and that the nonrelativistic approximation should be good. 
Therefore, first of all, the angular 
distribution of the tensor dipion momentum should be flat 
in the $X$ rest frame:
\begin{equation}
     d\Gamma/d\cos\theta_{\pi\pi} \simeq (1/2)\Gamma_0,
                     \label{flat}
\end{equation}
where the sign of $\simeq$ means ``in the nonrelativistic 
approximation''. This may not be an easy test since the fat 
dipions do not move much in the $X$ rest frame so that
determination of the ${\bf p}_{\pi\pi}$ direction is subject to 
large errors.

Eq. (\ref{flat}) only tests $L=0$, not directly 
the quantum numbers of the dipion. We shall be able to test 
the tensor nature of dipion by measuring the 
angular correlation of ${\bf p}_{\pi}= {\bf p}_{\pi^+}
- {\bf p}_{\pi^-}$ of the dipion rest frame with the 
$X$ momentum in the $B^{\pm}$ rest frame. 
Since angular momenta conservation holds among spins alone for 
$L=0$, the spin components of $\pi^+\pi^-$ and $J/\psi$ 
make up $|J,M\rangle = |2,0\rangle$ of $X(3872)$ as
\begin{equation}
    |2,0\rangle_{X}= (1/\sqrt{2})(|2,+1\rangle_{\pi\pi}
           |1,-1\rangle_{J/\psi}-|2,-1\rangle_{\pi\pi}
           |1,+1\rangle_{J/\psi}), \label{CG}
\end{equation}
when the spin quantization axis is chosen along the $X$
momentum in the $B^{\pm}$ rest frame. The $s_z=0$ states
are missing in the right-hand of Eq. (\ref{CG}), which is
characteristic of the Clebsch-Gordan composition involving 
$|j,m\rangle =|1,0\rangle$. This leads to the angular 
distribution of ${\bf p}_{\pi}$ as
\begin{equation}
    d\Gamma/d\cos\theta_{\pi} \simeq
 \frac{15}{4}\Gamma_0\cos^2\theta_{\pi}\sin^2\theta_{\pi},
        \;\;(J^{PC}=2^{--}) \label{AC1}
\end{equation}
where $\theta_{\pi}$ is the polar angle of ${\bf p}_{\pi}=
{\bf p}_{\pi^+} -{\bf p}_{\pi^-}$ in the dipion rest frame as 
measured from the $X$ momentum in the $B$ rest frame. 
Eq. (\ref{AC1}) can 
be obtained by an elementary calculation of the nonrelativistic decay 
amplitude made of three polarizations, $\epsilon^{jkl}
\varepsilon^{ij}_X\varepsilon^{ik}_{\pi\pi}
\varepsilon^{l}_{J/\psi}$. The test will be done most accurately 
if one selects the dipions of 700 MeV $< m_{\pi\pi} <$
775 MeV. The Belle data show many events
in this mass region relative to the region below 
it\cite{Belle1,Belle2}.

\subsection{Vector dipion}

The spin parity of  interest in the case of the vector dipion is  
$J^{PC}=1^{++}$  for the molecule.  
The vector dipion most often forms a $\rho$ meson. Since $\rho$ 
and $J/\psi$ are nearly at rest in the X rest frame, the 
relative orbital angular momentum between them should be $L=0$
just as in the case of the tensor dipion. 
In this circumstance the same simplification occurs. 
The angular distribution of ${\bf p}_{\pi\pi}$ 
tests $L=0$ by Eq. (\ref{flat}). The vector nature of the dipion 
can be tested by the angular correlation,
\begin{equation} 
    d\Gamma/d\cos\theta_{\pi} \simeq
              \frac{3}{4}\Gamma_0 \sin^2\theta_{\pi},
                 \;\; (J^{PC}=1^{++}), \label{AC2}
\end{equation}
where $\theta_{\pi}$ is defined in the same way as in 
Eq. (\ref{AC1}). 

\section{Summary}

   The most likely candidate for the narrow resonance discovered 
by the Belle Collaboration is the $1^3D_2$ charmonium state. 
Depending on the outcome of the radiative decay measurement, 
however, the $2^1P_1$ charmonium may become a better alternative. 
Among the molecules, the $J^{PC}=1^{+\pm}$ molecule states of 
$D\overline{D}^*+\overline{D}D^*$ are acceptable as far as quantum
numbers are concerned and furthermore would easily satisfy the 
constraints imposed by the absence of radiative decay mode.
On the other hand, the final states of $\pi^+\pi^-J/\psi$ with a 
scalar dipion and
a vector dipion cannot be produced by a single resonance because of
charge conjugation invariance no matter how badly 
isospin symmetry is violated. 

The crucial information leading to 
determination of the quantum numbers of this resonance should come
from spin of the dipion. The various angular distributions and 
correlations that have been presented here will help us in 
identifying the nature of this resonant state conclusively.

\acknowledgements

We thank S. Olsen for continuing encouragement.
One of the authors (MS) thanks J. D. Jackson for discussion on the
angular distribution.
This work was supported by the Director, Office of Science, 
Office of High Energy and Nuclear Physics, Division of High Energy Physics,
of the U.S. Department of Energy under contract DE--AC03--76SF00098 
and DE--FG03--94ER40833, and
in part by the National Science Foundation under grant PHY-0098840.

\begin{table}
\caption{Unnatural spin-parity and charge-psrity state. $L$ stands for 
the relative orbital angular momentum of $\pi^+\pi^-$ and $J/\psi$.}
\begin{tabular}{llll} 
$J_{\pi\pi}^{PC}$ & I &  L & $J^{PC}$  \\ \hline
$0^{++}$   &   0, 2  &  1 & $0^{+-}, 1^{+-}, 2^{+-}$  \\
           &          &  2 & $2^{--}$ \\
$2^{++}$   &   0, 2  &  0 & $2^{--}$ \\
           &          &  1 & $0^{+-}, 1^{+-}, 2^{+-}, 3^{+-}, 4^{+-}$\\          
$1^{--}$   &   1      &  0 & $1^{++}$ \\ 
           &          &  1 & $0^{-+}, 1^{-+}, 2^{-+}, 3^{-+}$   
\end{tabular}
\label{table:1}
\end{table}


\begin{references}

\bibitem{Belle1} Belle Collaboration, K. Abe et al., hep-ex/0308029.
\bibitem{Belle2} Belle Collaboration, S. K. Choi et al., hep-ex/0309032. 
\bibitem{Eich} E.J. Eichten, K. Lane, and C. Quigg, Phys. Rev. Lett. 
            {\bf 89}, 162002 (2002). See also P. Ko, J. Lee, and H.S. Song, 
            Phys. Lett. {\bf B395}, 107 (1997).
\bibitem{Georgi} M.B. Voloshin and L.B. Okun, JETP Lett. {\bf 23}, 333
        (1976); A. De Rujula, H. Georgi, and S.L. Glashow, Phys. Rev.
         Lett. {\bf 38}, 317 (1977). See also N. T\"{o}rnqvist, Phys. 
         Rev. Lett. {\bf 67}, 556 (1991);
         T.E.O. Ericson and G. Karl, Phys. Lett. {\bf B309} 426 (1993).
\bibitem{Pot} E. Eichten, K. Gottfried, T. Kinoshita, K. Lane, and
         T.M. Yan, Phys. Rev. D{\bf 21}, 203 (1980); W. Buchm\"{u}ller 
        and S-H.H. Tye, Phys. Rev. D{\bf 24}, 132 (1981):
\bibitem{Isgur} S. Godfrey and N, Isgur, Phys. Rev. D{\bf 32}, 189 (1985).
\bibitem{Torn1}  N. T\"{o}rnqvist, hep-ph/0308277; F.E. Close and 
                P.R. Page, hep-ph/0309253.
\bibitem{Torn2} N. T\"{o}rnqvist, Z. Phys. {\bf C61}, 525,(1994).
\bibitem{B}  T.E. Browder, S. Pakvasa, and A.A. Petrov, hep-ph/0307054.
           The same mixing was proposed for $D_{sJ}(2317)$.
\bibitem{S} M. Suzuki, hep-ph/0307118.
\bibitem{JW} M. Jacob and J.C. Wick, Ann. Phys. {\bf 7}, 404 (1959).
\end{references}
\end{document}